\documentclass[%
reprint,
superscriptaddress,
amsmath,amssymb,
aps,longbibliography,
]{revtex4-2}

\usepackage{graphicx}%
\usepackage{verbatim}
\usepackage[version=4]{mhchem}
\usepackage{xcolor}
\usepackage{physics}
\usepackage[normalem]{ulem}
\usepackage{pdfpages}
\usepackage{pgffor}
\usepackage{booktabs}

\usepackage{listings}
\usepackage{xcolor}
\usepackage{tcolorbox}

\usepackage{listings}

\makeatletter
\AtBeginDocument{\let\LS@rot\@undefined}
\makeatother

\begin{document}
\title{\texttt{scicode-widgets}: Bringing Computational Experiments \\to the Classroom with Jupyter Widgets}

\author{Alexander Goscinski}
\affiliation{PSI Center for Scientific Computing, Theory and Data, 5232 Villigen PSI, Switzerland}

\author{Taylor J. Baird}
\affiliation{Centre Européen de Calcul Atomique et Moléculaire (CECAM), École Polytechnique Fédérale de Lausanne, 1015 Lausanne, Switzerland}

\author{Dou Du}
\affiliation{Centre Européen de Calcul Atomique et Moléculaire (CECAM), École Polytechnique Fédérale de Lausanne, 1015 Lausanne, Switzerland}
\affiliation{Theory and Simulation of Materials (THEOS), École Polytechnique Fédérale de Lausanne, 1015 Lausanne, Switzerland}

\author{João Prado}
\affiliation{Laboratory of Computational Science and Modeling, École Polytechnique Fédérale de Lausanne, 1015 Lausanne, Switzerland}

\author{Divya Suman}
\affiliation{Laboratory of Computational Science and Modeling, École Polytechnique Fédérale de Lausanne, 1015 Lausanne, Switzerland}

\author{Tulga-Erdene Sodjargal}
\affiliation{Laboratory of Computational Science and Modeling, École Polytechnique Fédérale de Lausanne, 1015 Lausanne, Switzerland}

\author{Sara Bonella}
\affiliation{Centre Européen de Calcul Atomique et Moléculaire (CECAM), École Polytechnique Fédérale de Lausanne, 1015 Lausanne, Switzerland}

\author{Giovanni Pizzi}
\affiliation{PSI Center for Scientific Computing, Theory and Data, 5232 Villigen PSI, Switzerland}
\affiliation{National Centre for Computational Design and Discovery of Novel Materials (MARVEL), 5232 Villigen PSI, Switzerland}

\author{Michele Ceriotti}
\affiliation{Laboratory of Computational Science and Modeling, École Polytechnique Fédérale de Lausanne, 1015 Lausanne, Switzerland}

\begin{abstract}
``Computational experiments'' use code and interactive visualizations to convey mathematical and physical concepts in an intuitive way, and are increasingly used to support \emph{ex cathedra} lecturing in scientific and engineering disciplines.
Jupyter notebooks are particularly well-suited to implement them, but involve large amounts of ancillary code to process data and generate illustrations, which can distract students from the core learning outcomes. 
For a more engaging learning experience that only exposes relevant code to students---allowing them to focus on the interplay between code, theory and physical insights---we developed \texttt{scicode-widgets} (released as \texttt{scwidgets}), a Python package to build Jupyter-based applications. 
The package facilitates the creation of interactive exercises and demonstrations for students in any discipline in science, technology and engineering.
Students are asked to provide pedagogically meaningful contributions in terms of theoretical understanding, coding ability, and analytical skills.
The library provides the tools to connect custom pre- and post-processing of students' code, which runs seamlessly ``behind the scenes'', with the ability to test and verify the solution, as well as to convert it into live interactive visualizations driven by Jupyter widgets.
\end{abstract}

\maketitle

Jupyter notebooks~\cite{jupyter} have been used extensively for the creation of educational contents, with applications in chemistry~\cite{notebooks-in-chemistry}, physics~\cite{notebooks-in-physics,urcelay-olabarria_jupyter_2017,tufino_using_2025} and mathematics~\cite{notebooks-in-mathematics}.
The cell-based structure of the notebook supports conveying the teaching material in smaller units that can be connected in a sequential coherent narrative, thereby facilitating learning~\cite{jupyter-narrative}.
In addition to bare notebooks, custom Jupyter widgets have been developed both in research contexts~\cite{aiidalab} and in educational contexts to create virtual lab applications, e.g., allowing students to explore the experimental parameter space interactively~\cite{sutchenkov2020active,du2023osscar,du2024widget} or simply serving as a replacement for a hands-on lab experience~\cite{vrl-survey}.
Indeed, studies in control education have observed that these learning experiences can reach a similar effectiveness as their hands-on counterparts~\cite{vrl-survey}.
These examples usually focus on conveying the interaction between theory and experimental results to students. However, they do not teach how to create and design these computational experiments, a skill that is becoming more essential for all fields of computational science.
Although related skills are taught in separate courses, the interplay of theoretical understanding, algorithmic design, and interpretation of the results is typically only taught in more advanced courses.
A crucial contributing factor to its late introduction in the curriculum is the amount of boilerplate code that needs to be written to process and visualize the data, together with the amount of time needed to get acquainted with the domain-specific tools that help with this kind of processing.
While the code for data processing can be provided by the teacher, it still dilutes students' focus if the notebook is populated by ancillary code that does not contribute to the core learning outcomes.
In addition, for an interactive experience, computational experiments need to be rerun frequently to assess the results under different conditions.
While parameters in a notebook cell can be changed and rerun, it still adds friction to the experience of students if multiple actions are required to rerun an experiment (such as changing values in several cells and rerunning them in a specific order).
Due to these aspects, learning is hindered by having to perform repetitive actions.
To address these challenges, we developed the Python library \texttt{scicode-widgets} (released as \texttt{scwidgets}) that facilitates the implementation of Jupyter-based web applications,
providing a seamlessly interactive learning experience of computational experiments to students.
\texttt{scicode-widgets} allows teachers to expose only the essential code, parameters, and results (such as interactive plots) to students, while hiding the unnecessary details in the background, thus helping them focus first on the core scientific and computational concepts.

\begin{figure}
    \centering \includegraphics[width=0.49\textwidth]{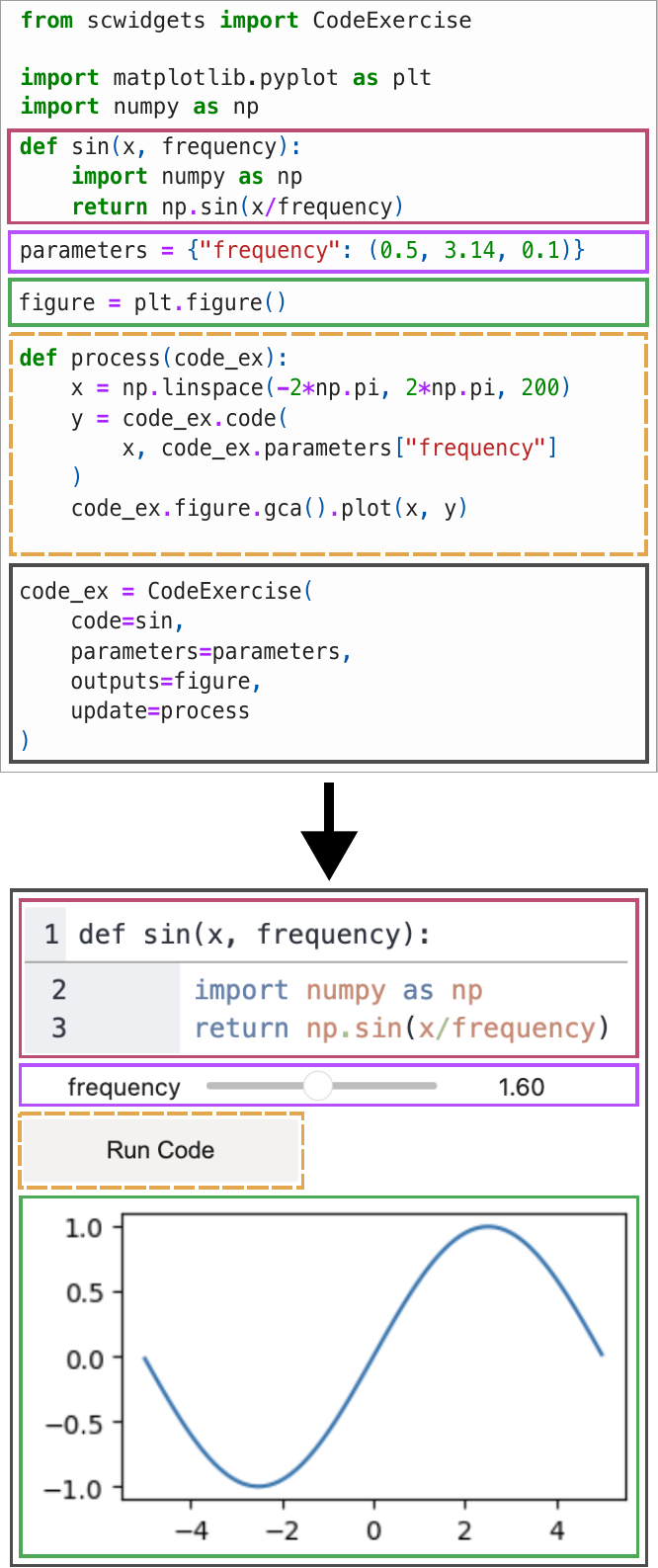}
    \caption{An example of how the coding environment, the parameter panel and the visualization of an interactive exercise widget are defined by the teacher (top) and how they are visualized by the corresponding widgets (bottom).}
    \label{fig:exercise-code}
\end{figure}

\section{COMPUTATIONAL EXPERIMENTS}
The main use case of \texttt{scicode-widgets} is the  creation of flexible code exercises or demonstrations for students in the form of an interactive application built by combining several Jupyter widgets.
This application can be divided into three parts that are marked with different colors in the bottom part of Fig.~\ref{fig:exercise-code}: A coding environment (red) in which students can implement the algorithm derived from the provided theory, a panel of parameters (purple) where students can manipulate the experimental settings, and a visualization (green) that is generated from the code of the students utilizing the experimental settings specified in the parameter panel.
To provide full flexibility in the creation of an exercise, the teacher implements a function (orange box in the upper panel of Fig.~\ref{fig:exercise-code}) that defines
the logic to process the parameters specified in the parameters panel, and to input these as arguments to the code supplied by the students, along with functionality to visualize the results of the code execution, e.g. in Fig.~\ref{fig:exercise-code} in the form of a plot.
This process function is then executed by a click of the ``Run code'' button (marked in orange in the bottom panel of Fig.~\ref{fig:exercise-code}).
Alternatively, the code exercise can be configured to run such process function upon the change of any of the parameters in the panel of parameters.
This simple example shows how the interactive view of a  notebook based on \texttt{scicode-widgets} hides away plotting and data processing code, generating a clean interface that helps the student focus on pedagogically meaningful elements. 

\begin{figure}
    \centering    \includegraphics[width=0.49\textwidth]{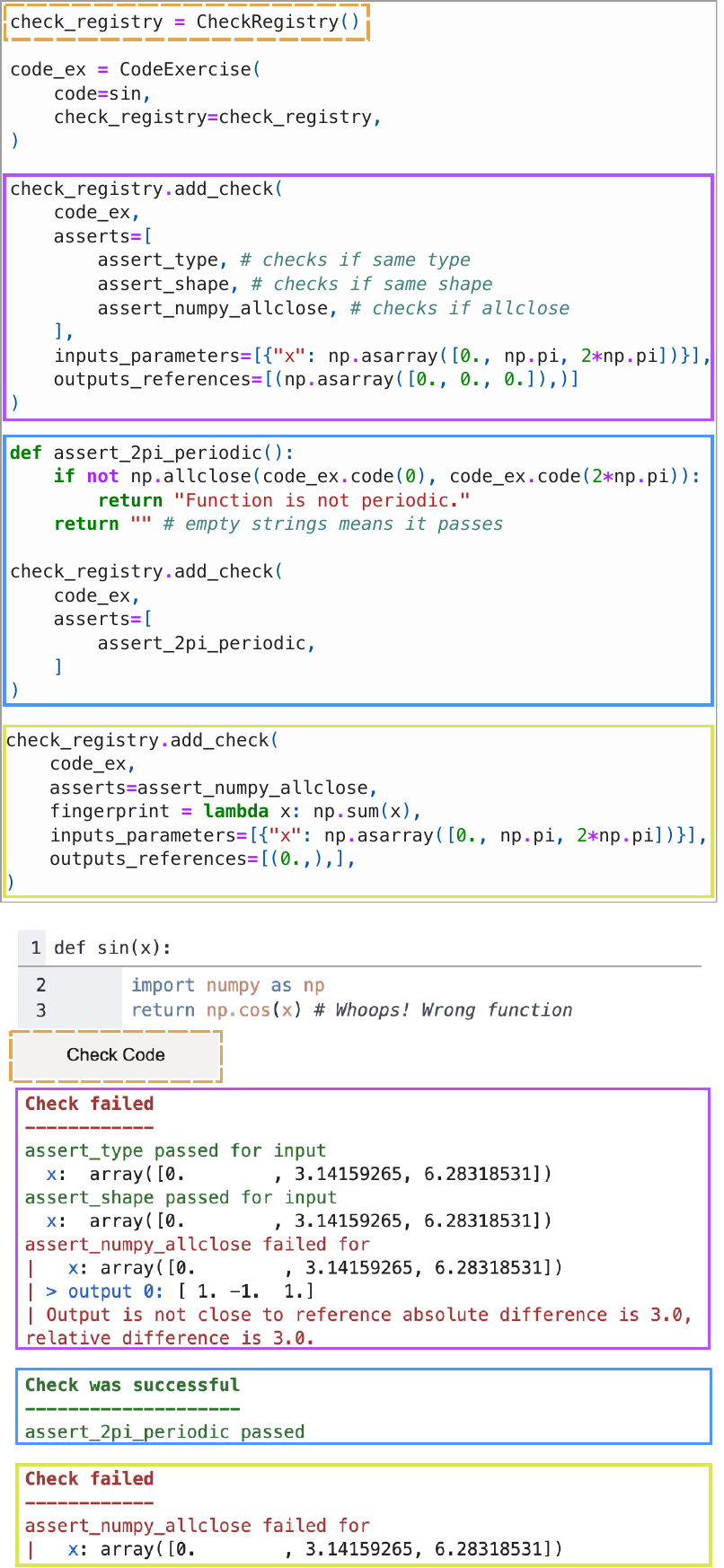}
    \caption{Examples of the check functionality. Each check is framed with the same color as its corresponding output. The purple check verifies the built-in Python type, shape and value closeness between the output of the students' code and some reference output provided by the teacher. The blue check verifies the periodicity of the function provided by the students, i.e., $\sin(0) = \sin(2\pi)$. The yellow-green check uses a fingerprint function, demonstrating how to prevent disclosing the reference output to the students. For instance, in this example only the sum of the values of the function on a specified grid of $x$ values (here only composed of three points for illustrative purposes) is checked against a reference value.}
    \label{fig:check-code}
\end{figure}

\subsection{Check solutions}
The instructor can also define checks for the code provided by the students, to support them with a quick feedback on the correctness of their answer, thus facilitating approaching the correct solution. An example is provided in Fig.~~\ref{fig:check-code}.
A variety of checks typically used in scientific environments are already implemented, including type and equality checks, as well as shape and numerical closeness checks for \texttt{numpy} arrays.
These checks work as unit tests, validating the output of the code of the students for a set of input and reference values provided by the teacher.
The instructor can also create checks that validate specific functional behavior, e.g. periodicity for trigonometric functions (for an example, see the check framed by a blue box in Fig.~\ref{fig:check-code}).
One noteworthy point is that, for certain exercises, the reference values can contain hints of the solution, which can be exploited by students to solve the exercise.
As the reference values need to be part of the notebook, they cannot be safely hidden from the students.
The library therefore provides an option to pass a one-way function, analogous to a hashing or fingerprint function, that post-processes the output before the check is run.
Since no inference can be made once the output is processed through a one-way function, the reference values of the checks do not reveal any information about the actual solution of the exercise.
This fingerprint function depends highly on the domain and therefore needs to be provided by the instructor. Nevertheless, the library exposes a simple interface to facilitate providing such checks.
An example of a check using the fingerprint functionality can be seen in Fig.~\ref{fig:check-code}, framed by a yellow-green box.

\begin{figure}
    \centering
    \includegraphics[width=0.5\textwidth]{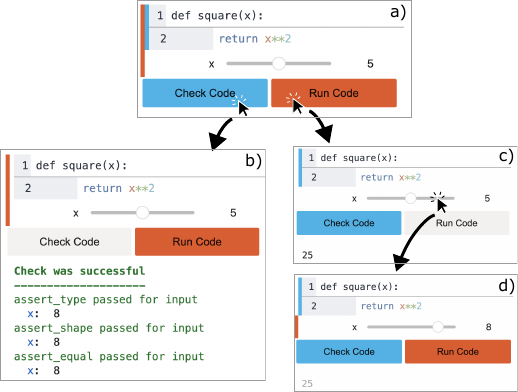}
    \caption{An example of how the cue system works, showing a sequence of widget interactions. Panel a) shows both blue and red cues active, resulting for instance from a previous change by the student of both the code and the value of the $x$ parameter, without yet having rerun neither the code (to update the visualization) nor the checks. From a) to b) the ``Check Code'' button is pressed, removing the blue cues from the button and the code input (but keeping the red cues, as the code has not been rerun yet, and thus any related visualization are not yet updated). From a) to c) the ``Run Code'' button is pressed, removing the red cues from the code input, the parameter panel and the button (but keeping the blue cues, as the checks have not been run yet). From c) to d) the parameter $x$ is changed, resulting in a new red cue both on the parameter panel and the button.}
    \label{fig:cue-system}
\end{figure}

\subsection{Cues for readability}
While students work with \texttt{scicode-widgets}, they might often change their provided answer without immediately rerunning the code (to update the visualizations) or performing the related checks.
This might leave the student unaware of which parts of the widget application are outdated, i.e., which parts need to be rerun to update the outputs.
We therefore implemented a cue system that highlights the parts of the widget application that were modified since the last update and thus need to be rerun.
When a student modifies an input in the widget, the system highlights the affected part with vertical bars of different colors -- referred to as a ``cues'' -- depending on the corresponding needed action: blue for checking, and red for running the code.
This indicates that the corresponding output is outdated and immediately visualizes which inputs have been changed.
An example of several interactions and their effect on the cues can be seen in Fig.~\ref{fig:cue-system}.
This real-time feedback ensures that students are always aware of the current state of the outputs and can understand more easily the effect of their modifications, resulting in a more dynamic learning process.

\section{Learning ridge regression through an interactive notebook}

\begin{figure}
    \centering    \includegraphics[width=0.5\textwidth]{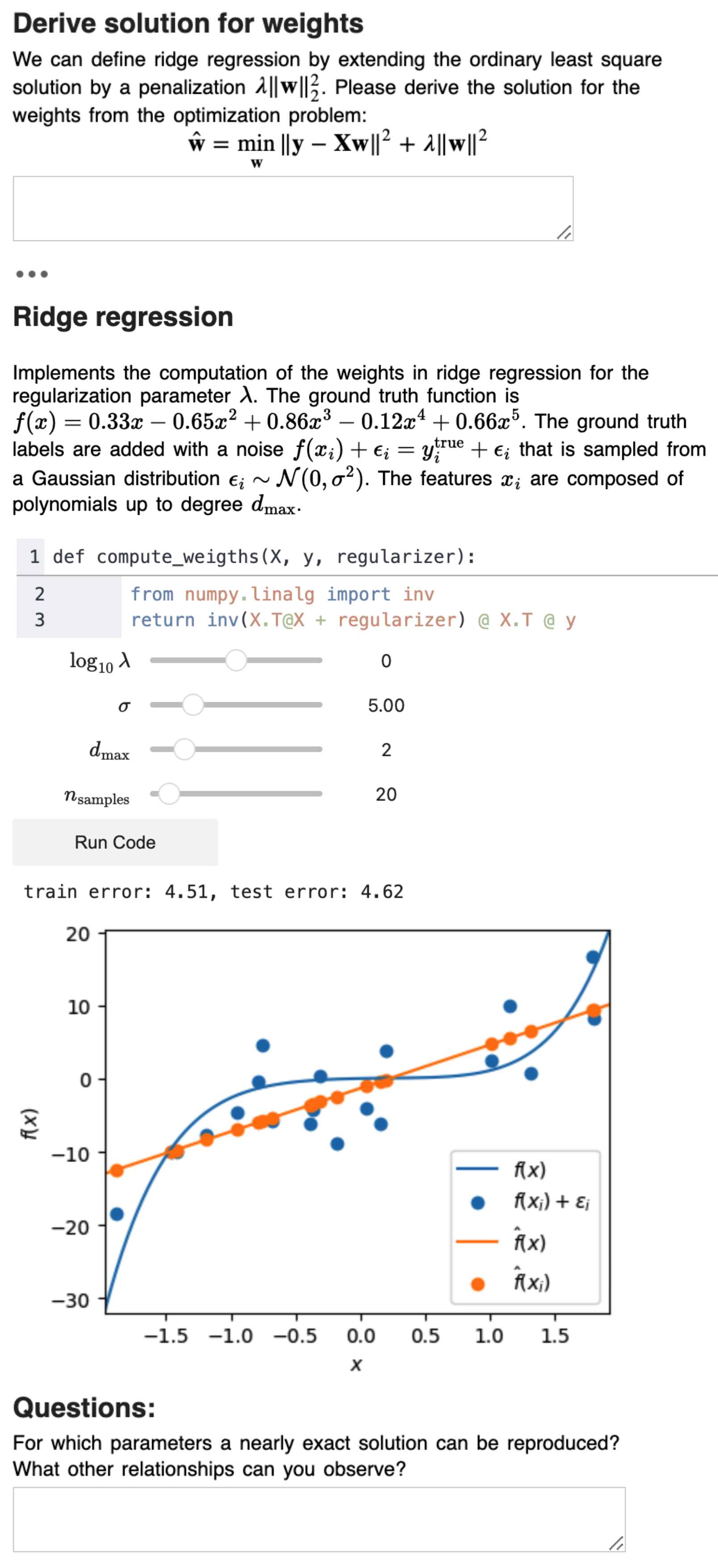}
    \caption{An example notebook covering the topic of ridge regression, divided into exercises demanding mathematical coding and analytical skills. It includes both our exercise widget and free-text boxes for answering questions. Even just generating this simple plot requires more than $30$ lines of Python code. Thanks to \texttt{scicode-widgets}, this code is hidden from students, so that they can instead focus on understanding the core concepts of ridge regression and only implement the core functions (in this case, they only need to provide the body of the \texttt{compute\_weights} Python function).}
    \label{fig:ridge}
\end{figure}

In this section, we showcase the use of \texttt{scicode-widgets} for an exercise aimed at explaining ridge regression~\cite{bishop} to students.
A notebook that implements such an exercise is shown in Fig.~\ref{fig:ridge}.
In the following subsections we briefly discuss ways in which the code exercise creates a more engaging learning experience with respect to more traditional learning approaches.

\subsection{Validate solutions and identify mistakes}
An interesting approach to use the application is to first ask students to derive a solution for the weights from the optimization problem defining ridge regression, and then to implement it in the code environment widget.
If the students make a mistake in the mathematical derivation or in the code implementation, they need to identify where the problem lies.
While the instructor could provide explicit checks so that students can validate their solution, a more engaging learning experience is created by providing hints on how to use the visualization for validation.
Students could then for instance reason that, when the noise and regularization strength are very small, the solution of ridge regression approaches the exact solution, which is clearly illustrated in the plot produced by the widget app.
The instructor could help focus students' attention by hinting at this fact with a question of the form ``For which parameters can a nearly exact solution be reproduced? Can you observe this with your code?'', thereby provoking the train of thought to use this as a validation.
Such an experience teaches students to find mistakes in complex processes, requiring them to systematically exclude parts by validating their correctness and identify the parts that are relevant for investigation.

\subsection{Observe relationships through interactivity}
After implementing the code, students can directly proceed with experimental observations.
The plot in Fig.~\ref{fig:ridge} shows the ground truth function $f$, the data points with noise $f(x_i) + \epsilon_i$, and the predicted function $\hat{f}(x)$.
First, the possibility to manipulate every parameter (via the sliders above the plot) effectively communicates to students the impact of each parameter on a visual level.
Second, through the manipulation of the parameters, students can analyze their relationship, e.g., how the regularization strength $\lambda$ and the number of data points $n_\text{samples}$ affect the test error for different degrees of the polynomial function $d_\text{max}$ used for fitting.
While this is typically a challenging task, as it requires exploring the interplay of four parameters, the corresponding plot provides students with direct feedback (and the instructor can further hint at certain relationships in subsequent questions).

\section{Technical challenges and solutions}
\texttt{scicode-widgets} depends on several external libraries that implement the widgets used as components in the applications.
In addition to the package \texttt{ipywidgets}, that provides most of the standard widgets used for the applications, the coding environment relies on our package \texttt{widget-code-input}, providing the editable text box where students can input their code, where the function signature (and possibly function documentation in the format of a standard Python "docstring") are hardcoded and read-only (see its appearance in Figs.~\ref{fig:exercise-code}, \ref{fig:check-code}, \ref{fig:cue-system} and \ref{fig:ridge}). While this widget might seem redundant, as notebook cells already allow students to edit code, it becomes essential when our widgets are combined with appmode~\cite{appmode} or similar Jupyter extensions that run a complete notebook hiding all input cells, thus providing an ``app-like'' appearance to the notebook. Indeed, even if \texttt{scicode-widgets} reduces to a minimum the need for boilerplate code (see the top part of Fig.~\ref{fig:exercise-code}), hiding completely all code and showing only the resulting widgets further helps students in focusing only on the learning goal (see bottom of Fig.~\ref{fig:exercise-code}).
Apart from some custom style sheets (CSS) code for styling, \texttt{scicode-widgets} does not include any JavaScript code and builds its widget applications from the building blocks provided by its dependencies.
In comparison to a two-language solution that would use JavaScript for the frontend combined with a Python interface for the code, this design choice has the advantage that the library is easier to maintain, as well as being accessible to a broader range of teaching instructors for code adaption, given Python's position as one of the most commonly used languages in an academic, education and research setting. 
In the following, we further discuss in more detail some of the technical challenges that we encountered during the development and the solutions we implemented.

\subsection{Cued widgets}
The exercise widget app is modular, allowing flexible initialization with only the components needed for a given use case. For example, the parameter box can be disabled in fixed experimental settings.
The initialization process needs to handle several cases, which complicates the logic.
To reduce the code redundancy, the logic for the \text{cue} system is handled in dedicated ``cue widgets'',  that activate the appropriate visual cue when it observes a change in one of the traits~\cite{traitlets} in a list of widgets provided on initialization (i.e., in any of the other widgets that it observes for change).
For the majority of the cueing, we use a \texttt{cue box} that wraps around an existing widget to highlight it around the edges.
The behavioral change when a widget is cued is defined in a CSS file to allow for customization by the teaching instructor (e.g., color or line thickness).
Furthermore, for resetting the cue, we implement a \texttt{reset cue} button that only resets the cue on a successful action.
Moving this code logic into the individual components simplified handling of the logic in the exercise widget application and testing their functionality and correctness.

\subsection{Saving mechanism}
Since the target of our package is creating exercises for students, who then typically need to hand in their solution for grading, we need a mechanism to save the student solutions and transfer them to the teaching instructor.
Jupyter environments can store widget states, including the HTML and JavaScript part of the widget.
The execution of the code widget, however, requires communication with a Python kernel that runs the functions required for the data processing.
We therefore need to re-execute the initialization of the widgets to recreate the corresponding Python objects, which would result in overwriting the static widget state, including the solutions provided by the students.
To solve this issue, we implemented a custom saving mechanism for the exercise widgets, which stores both the code and the parameter values in a JavaScript Object Notation (JSON) file.
The logic responsible for storing and loading the answers is in the exercise manager Python class, which retrieves the answer of an exercise widget. 

\subsection{Testing of the user interface}
For an effective teaching experience, the robustness of the exercise application is essential.
Indeed, the students are already required to consider errors in their understanding of the theory, the code, and interpretation of results.
Adding failure of the widgets on top of these hurdles would affect very negatively the teaching experience.
We therefore test the user interface of the exercise application extensively within a unit test framework, using the package \texttt{selenium}~\cite{selenium}, which allows us to simulate realistic user interactions (e.g. typing on the keyboard or clicking with the mouse) via the web browser, and test the graphical output of the application.

\section{CONCLUSION}
In this work, we presented how an interactive widget application that combines a coding environment and visualization can be used to provide an engaging learning experience for students.
The creation of the widget application leaves enough room for flexibility to implement arbitrary logic for processing the student code, while taking care of the widget logic for the teaching instructor.
This allows the teaching instructor to let arbitrary pre- and postprocessing of the student code happen ``behind the scenes'', so that students can fully focus on the main learning goal.
We demonstrated the usage of \texttt{scicode-widgets} for teaching ridge regression, showing how the interactivity of the widgets enables seamless transitions between theory, code and analysis, resulting in interactions that would not occur as easily without them.
\texttt{scicode-widgets} supports the most recent JupyterLab \cite{jupyterlab} version (version 4 at the point of writing) and can be easily installed with a \texttt{pip} command.
We therefore expect it to be straightforward to deploy and integrate into any local or cloud-based Jupyter deployment.
These widgets have already been used to prepare a course at EPFL (MSE-305, Introduction to Atomic-Scale Modeling), and were refined based on the feedback from the students.

\section{DATA AVAILABILITY}
The code producing the \texttt{scicode-widgets} applications underlying each figure is available on Zenodo~\cite{supplementary-code}.
The notebooks used in the MSE-305 course are distributed in a dedicated repository~\cite{iam-notebooks}.

\section{ACKNOWLEDGMENTS}

We acknowledge financial support from the EPFL Open Science Fund
via the OSSCAR project, from the EPFL DRIL fund titled ``Jupyter web applications for quantum simulations'', and from the NCCR MARVEL, a National Centre of Competence in Research, funded by the Swiss National Science Foundation (grant number 205602).
We acknowledge CECAM for dedicated OSSCAR dissemination activities.
A.G. and G.P. acknowledge funding by the SwissTwins project, funded by the Swiss State Secretariat for Education, Research and Innovation (SERI).
M.C. and D.S. acknowledge funding from the European Research Council (ERC) under the research and innovation program (Grant Agreement No. 101001890-FIAMMA). 
We acknowledge useful discussions and feedback from Cécile Hardebolle and the students of the EPFL MSE-305 course -- especially those of the academic year 2021-2022.


\begin{thebibliography}{19}%
\makeatletter
\providecommand \@ifxundefined [1]{%
 \@ifx{#1\undefined}
}%
\providecommand \@ifnum [1]{%
 \ifnum #1\expandafter \@firstoftwo
 \else \expandafter \@secondoftwo
 \fi
}%
\providecommand \@ifx [1]{%
 \ifx #1\expandafter \@firstoftwo
 \else \expandafter \@secondoftwo
 \fi
}%
\providecommand \natexlab [1]{#1}%
\providecommand \enquote  [1]{``#1''}%
\providecommand \bibnamefont  [1]{#1}%
\providecommand \bibfnamefont [1]{#1}%
\providecommand \citenamefont [1]{#1}%
\providecommand \href@noop [0]{\@secondoftwo}%
\providecommand \href [0]{\begingroup \@sanitize@url \@href}%
\providecommand \@href[1]{\@@startlink{#1}\@@href}%
\providecommand \@@href[1]{\endgroup#1\@@endlink}%
\providecommand \@sanitize@url [0]{\catcode `\\12\catcode `\$12\catcode
  `\&12\catcode `\#12\catcode `\^12\catcode `\_12\catcode `\%12\relax}%
\providecommand \@@startlink[1]{}%
\providecommand \@@endlink[0]{}%
\providecommand \url  [0]{\begingroup\@sanitize@url \@url }%
\providecommand \@url [1]{\endgroup\@href {#1}{\urlprefix }}%
\providecommand \urlprefix  [0]{URL }%
\providecommand \Eprint [0]{\href }%
\providecommand \doibase [0]{https://doi.org/}%
\providecommand \selectlanguage [0]{\@gobble}%
\providecommand \bibinfo  [0]{\@secondoftwo}%
\providecommand \bibfield  [0]{\@secondoftwo}%
\providecommand \translation [1]{[#1]}%
\providecommand \BibitemOpen [0]{}%
\providecommand \bibitemStop [0]{}%
\providecommand \bibitemNoStop [0]{.\EOS\space}%
\providecommand \EOS [0]{\spacefactor3000\relax}%
\providecommand \BibitemShut  [1]{\csname bibitem#1\endcsname}%
\let\auto@bib@innerbib\@empty
\bibitem [{\citenamefont {Thomas}\ \emph {et~al.}(2016)\citenamefont {Thomas},
  \citenamefont {Benjamin}, \citenamefont {Fernando}, \citenamefont {Brian},
  \citenamefont {Matthias}, \citenamefont {Jonathan}, \citenamefont {Kyle},
  \citenamefont {Jessica}, \citenamefont {Jason}, \citenamefont {Sylvain},
  \citenamefont {Paul}, \citenamefont {Dami\'an}, \citenamefont {Safia},
  \citenamefont {Carol},\ and\ \citenamefont {Team}}]{jupyter}%
  \BibitemOpen
  \bibfield  {author} {\bibinfo {author} {\bibfnamefont {K.}~\bibnamefont
  {Thomas}}, \bibinfo {author} {\bibfnamefont {R.-K.}\ \bibnamefont
  {Benjamin}}, \bibinfo {author} {\bibfnamefont {P.}~\bibnamefont {Fernando}},
  \bibinfo {author} {\bibfnamefont {G.}~\bibnamefont {Brian}}, \bibinfo
  {author} {\bibfnamefont {B.}~\bibnamefont {Matthias}}, \bibinfo {author}
  {\bibfnamefont {F.}~\bibnamefont {Jonathan}}, \bibinfo {author}
  {\bibfnamefont {K.}~\bibnamefont {Kyle}}, \bibinfo {author} {\bibfnamefont
  {H.}~\bibnamefont {Jessica}}, \bibinfo {author} {\bibfnamefont
  {G.}~\bibnamefont {Jason}}, \bibinfo {author} {\bibfnamefont
  {C.}~\bibnamefont {Sylvain}}, \bibinfo {author} {\bibfnamefont
  {I.}~\bibnamefont {Paul}}, \bibinfo {author} {\bibfnamefont {A.}~\bibnamefont
  {Dami\'an}}, \bibinfo {author} {\bibfnamefont {A.}~\bibnamefont {Safia}},
  \bibinfo {author} {\bibfnamefont {W.}~\bibnamefont {Carol}},\ and\ \bibinfo
  {author} {\bibfnamefont {J.~D.}\ \bibnamefont {Team}},\ }\bibfield  {title}
  {\bibinfo {title} {Jupyter notebooks -- a publishing format for reproducible
  computational workflows},\ }in\ \href
  {https://doi.org/10.3233/978-1-61499-649-1-87} {\emph {\bibinfo {booktitle}
  {Positioning and Power in Academic Publishing: Players, Agents and
  Agendas}}}\ (\bibinfo  {publisher} {IOS Press},\ \bibinfo {year} {2016})\
  pp.\ \bibinfo {pages} {87--90}\BibitemShut {NoStop}%
\bibitem [{\citenamefont {Stokes}(2021)}]{notebooks-in-chemistry}%
  \BibitemOpen
  \bibfield  {author} {\bibinfo {author} {\bibfnamefont {G.~Y.}\ \bibnamefont
  {Stokes}},\ }\bibfield  {title} {\bibinfo {title} {How faculty with minimal
  programming experience implemented jupyter notebooks in physical and general
  chemistry courses},\ }in\ \href {https://doi.org/10.1021/bk-2021-1387.ch002}
  {\emph {\bibinfo {booktitle} {Teaching Programming across the Chemistry
  Curriculum}}}\ (\bibinfo  {publisher} {American Chemical Society},\ \bibinfo
  {year} {2021})\ Chap.~\bibinfo {chapter} {2}, pp.\ \bibinfo {pages}
  {13--27}\BibitemShut {NoStop}%
\bibitem [{\citenamefont {Hu}\ \emph {et~al.}(2021)\citenamefont {Hu},
  \citenamefont {Ahn}, \citenamefont {Lakatos}, \citenamefont {Bello},
  \citenamefont {McTague},\ and\ \citenamefont {Foley}}]{notebooks-in-physics}%
  \BibitemOpen
  \bibfield  {author} {\bibinfo {author} {\bibfnamefont {D.}~\bibnamefont
  {Hu}}, \bibinfo {author} {\bibfnamefont {J.~N.}\ \bibnamefont {Ahn}},
  \bibinfo {author} {\bibfnamefont {A.}~\bibnamefont {Lakatos}}, \bibinfo
  {author} {\bibfnamefont {J.}~\bibnamefont {Bello}}, \bibinfo {author}
  {\bibfnamefont {J.}~\bibnamefont {McTague}},\ and\ \bibinfo {author}
  {\bibfnamefont {J.~J.~I.}\ \bibnamefont {Foley}},\ }\bibfield  {title}
  {\bibinfo {title} {Integrating programming to reinforce quantum mechanical
  principles in physical chemistry}\ }(\bibinfo  {publisher} {American Chemical
  Society},\ \bibinfo {year} {2021})\ Chap.~\bibinfo {chapter} {7}, pp.\
  \bibinfo {pages} {89--105}\BibitemShut {NoStop}%
\bibitem [{\citenamefont {Urcelay-Olabarria}\ \emph {et~al.}()\citenamefont
  {Urcelay-Olabarria}, \citenamefont {Lazkoz}, \citenamefont {Urrestilla},
  \citenamefont {Leonardo},\ and\ \citenamefont
  {Igartua}}]{urcelay-olabarria_jupyter_2017}%
  \BibitemOpen
  \bibfield  {author} {\bibinfo {author} {\bibfnamefont {I.}~\bibnamefont
  {Urcelay-Olabarria}}, \bibinfo {author} {\bibfnamefont {R.}~\bibnamefont
  {Lazkoz}}, \bibinfo {author} {\bibfnamefont {J.}~\bibnamefont {Urrestilla}},
  \bibinfo {author} {\bibfnamefont {A.}~\bibnamefont {Leonardo}},\ and\
  \bibinfo {author} {\bibfnamefont {J.~M.}\ \bibnamefont {Igartua}},\
  }\bibfield  {title} {\bibinfo {title} {Jupyter notebook as the physics
  experimental laboratory's logbook - first approach:},\ }in\ \href
  {https://doi.org/10.5220/0006352104580463} {\emph {\bibinfo {booktitle}
  {Proceedings of the 9th International Conference on Computer Supported
  Education}}}\ (\bibinfo  {publisher} {{SCITEPRESS} - Science and Technology
  Publications})\ pp.\ \bibinfo {pages} {458--463}\BibitemShut {NoStop}%
\bibitem [{\citenamefont {Tufino}\ \emph {et~al.}()\citenamefont {Tufino},
  \citenamefont {Oss},\ and\ \citenamefont {Alemani}}]{tufino_using_2025}%
  \BibitemOpen
  \bibfield  {author} {\bibinfo {author} {\bibfnamefont {E.}~\bibnamefont
  {Tufino}}, \bibinfo {author} {\bibfnamefont {S.}~\bibnamefont {Oss}},\ and\
  \bibinfo {author} {\bibfnamefont {M.}~\bibnamefont {Alemani}},\ }\bibfield
  {title} {\bibinfo {title} {Using jupyter notebooks to foster computational
  skills and professional practice in an introductory physics lab course},\
  }\href {https://doi.org/10.1088/1742-6596/2950/1/012022} {\bibfield
  {journal} {\bibinfo  {journal} {Journal of Physics: Conference Series}\
  }\textbf {\bibinfo {volume} {2950}},\ \bibinfo {pages} {012022}}\BibitemShut
  {NoStop}%
\bibitem [{\citenamefont {Sümmermann}\ \emph {et~al.}(2021)\citenamefont
  {Sümmermann}, \citenamefont {Sommerhoff},\ and\ \citenamefont
  {Rott}}]{notebooks-in-mathematics}%
  \BibitemOpen
  \bibfield  {author} {\bibinfo {author} {\bibfnamefont {M.~L.}\ \bibnamefont
  {Sümmermann}}, \bibinfo {author} {\bibfnamefont {D.}~\bibnamefont
  {Sommerhoff}},\ and\ \bibinfo {author} {\bibfnamefont {B.}~\bibnamefont
  {Rott}},\ }\bibfield  {title} {\bibinfo {title} {Mathematics in the digital
  age: The case of simulation-based proofs},\ }\href
  {https://doi.org/10.1007/s40753-020-00125-6} {\bibfield  {journal} {\bibinfo
  {journal} {International Journal of Research in Undergraduate Mathematics
  Education}\ }\textbf {\bibinfo {volume} {7}},\ \bibinfo {pages} {438}
  (\bibinfo {year} {2021})}\BibitemShut {NoStop}%
\bibitem [{\citenamefont {Granger}\ and\ \citenamefont
  {Perez}(2021)}]{jupyter-narrative}%
  \BibitemOpen
  \bibfield  {author} {\bibinfo {author} {\bibfnamefont {B.~E.}\ \bibnamefont
  {Granger}}\ and\ \bibinfo {author} {\bibfnamefont {F.}~\bibnamefont
  {Perez}},\ }\bibfield  {title} {\bibinfo {title} {Jupyter: Thinking and
  storytelling with code and data},\ }\href
  {https://doi.org/10.1109/mcse.2021.3059263} {\bibfield  {journal} {\bibinfo
  {journal} {Computing in Science \& Engineering}\ }\textbf {\bibinfo {volume}
  {23}},\ \bibinfo {pages} {7–14} (\bibinfo {year} {2021})}\BibitemShut
  {NoStop}%
\bibitem [{\citenamefont {Yakutovich}\ \emph {et~al.}(2021)\citenamefont
  {Yakutovich}, \citenamefont {Eimre}, \citenamefont {Sch\"{u}tt},
  \citenamefont {Talirz}, \citenamefont {Adorf}, \citenamefont {Andersen},
  \citenamefont {Ditler}, \citenamefont {Du}, \citenamefont {Passerone},
  \citenamefont {Smit}, \citenamefont {Marzari}, \citenamefont {Pizzi},\ and\
  \citenamefont {Pignedoli}}]{aiidalab}%
  \BibitemOpen
  \bibfield  {author} {\bibinfo {author} {\bibfnamefont {A.~V.}\ \bibnamefont
  {Yakutovich}}, \bibinfo {author} {\bibfnamefont {K.}~\bibnamefont {Eimre}},
  \bibinfo {author} {\bibfnamefont {O.}~\bibnamefont {Sch\"{u}tt}}, \bibinfo
  {author} {\bibfnamefont {L.}~\bibnamefont {Talirz}}, \bibinfo {author}
  {\bibfnamefont {C.~S.}\ \bibnamefont {Adorf}}, \bibinfo {author}
  {\bibfnamefont {C.~W.}\ \bibnamefont {Andersen}}, \bibinfo {author}
  {\bibfnamefont {E.}~\bibnamefont {Ditler}}, \bibinfo {author} {\bibfnamefont
  {D.}~\bibnamefont {Du}}, \bibinfo {author} {\bibfnamefont {D.}~\bibnamefont
  {Passerone}}, \bibinfo {author} {\bibfnamefont {B.}~\bibnamefont {Smit}},
  \bibinfo {author} {\bibfnamefont {N.}~\bibnamefont {Marzari}}, \bibinfo
  {author} {\bibfnamefont {G.}~\bibnamefont {Pizzi}},\ and\ \bibinfo {author}
  {\bibfnamefont {C.~A.}\ \bibnamefont {Pignedoli}},\ }\bibfield  {title}
  {\bibinfo {title} {Aiidalab – an ecosystem for developing, executing, and
  sharing scientific workflows},\ }\href
  {https://doi.org/10.1016/j.commatsci.2020.110165} {\bibfield  {journal}
  {\bibinfo  {journal} {Computational Materials Science}\ }\textbf {\bibinfo
  {volume} {188}},\ \bibinfo {pages} {110165} (\bibinfo {year}
  {2021})}\BibitemShut {NoStop}%
\bibitem [{\citenamefont {Sutchenkov}\ and\ \citenamefont
  {Tikhonov}(2020)}]{sutchenkov2020active}%
  \BibitemOpen
  \bibfield  {author} {\bibinfo {author} {\bibfnamefont {A.~A.}\ \bibnamefont
  {Sutchenkov}}\ and\ \bibinfo {author} {\bibfnamefont {A.~I.}\ \bibnamefont
  {Tikhonov}},\ }\bibfield  {title} {\bibinfo {title} {Active investigation and
  publishing of calculation web based applications for studying process},\
  }\href {https://doi.org/10.1088/1742-6596/1691/1/012096} {\bibfield
  {journal} {\bibinfo  {journal} {Journal of Physics: Conference Series}\
  }\textbf {\bibinfo {volume} {1691}},\ \bibinfo {pages} {012096} (\bibinfo
  {year} {2020})}\BibitemShut {NoStop}%
\bibitem [{\citenamefont {Du}\ \emph {et~al.}(2023)\citenamefont {Du},
  \citenamefont {Baird}, \citenamefont {Bonella},\ and\ \citenamefont
  {Pizzi}}]{du2023osscar}%
  \BibitemOpen
  \bibfield  {author} {\bibinfo {author} {\bibfnamefont {D.}~\bibnamefont
  {Du}}, \bibinfo {author} {\bibfnamefont {T.~J.}\ \bibnamefont {Baird}},
  \bibinfo {author} {\bibfnamefont {S.}~\bibnamefont {Bonella}},\ and\ \bibinfo
  {author} {\bibfnamefont {G.}~\bibnamefont {Pizzi}},\ }\bibfield  {title}
  {\bibinfo {title} {{OSSCAR}, an open platform for collaborative development
  of computational tools for education in science},\ }\href@noop {} {\bibfield
  {journal} {\bibinfo  {journal} {Computer Physics Communications}\ }\textbf
  {\bibinfo {volume} {282}},\ \bibinfo {pages} {108546} (\bibinfo {year}
  {2023})}\BibitemShut {NoStop}%
\bibitem [{\citenamefont {Du}\ \emph {et~al.}(2024)\citenamefont {Du},
  \citenamefont {Baird}, \citenamefont {Eimre}, \citenamefont {Bonella},\ and\
  \citenamefont {Pizzi}}]{du2024widget}%
  \BibitemOpen
  \bibfield  {author} {\bibinfo {author} {\bibfnamefont {D.}~\bibnamefont
  {Du}}, \bibinfo {author} {\bibfnamefont {T.~J.}\ \bibnamefont {Baird}},
  \bibinfo {author} {\bibfnamefont {K.}~\bibnamefont {Eimre}}, \bibinfo
  {author} {\bibfnamefont {S.}~\bibnamefont {Bonella}},\ and\ \bibinfo {author}
  {\bibfnamefont {G.}~\bibnamefont {Pizzi}},\ }\bibfield  {title} {\bibinfo
  {title} {Jupyter widgets and extensions for education and research in
  computational physics and chemistry},\ }\href@noop {} {\bibfield  {journal}
  {\bibinfo  {journal} {Computer Physics Communications}\ }\textbf {\bibinfo
  {volume} {305}},\ \bibinfo {pages} {109353} (\bibinfo {year}
  {2024})}\BibitemShut {NoStop}%
\bibitem [{\citenamefont {Heradio}\ \emph {et~al.}(2016)\citenamefont
  {Heradio}, \citenamefont {de~la Torre},\ and\ \citenamefont
  {Dormido}}]{vrl-survey}%
  \BibitemOpen
  \bibfield  {author} {\bibinfo {author} {\bibfnamefont {R.}~\bibnamefont
  {Heradio}}, \bibinfo {author} {\bibfnamefont {L.}~\bibnamefont {de~la
  Torre}},\ and\ \bibinfo {author} {\bibfnamefont {S.}~\bibnamefont
  {Dormido}},\ }\bibfield  {title} {\bibinfo {title} {Virtual and remote labs
  in control education: A survey},\ }\href
  {https://doi.org/10.1016/j.arcontrol.2016.08.001} {\bibfield  {journal}
  {\bibinfo  {journal} {Computers \& Education}\ }\textbf {\bibinfo {volume}
  {98}},\ \bibinfo {pages} {14} (\bibinfo {year} {2016})}\BibitemShut {NoStop}%
\bibitem [{\citenamefont {Bishop}(2006)}]{bishop}%
  \BibitemOpen
  \bibfield  {author} {\bibinfo {author} {\bibfnamefont {C.~M.}\ \bibnamefont
  {Bishop}},\ }\href {https://doi.org/10.5555/1162264} {\emph {\bibinfo {title}
  {Pattern Recognition and Machine Learning (Information Science and
  Statistics)}}}\ (\bibinfo  {publisher} {Springer-Verlag},\ \bibinfo {address}
  {Berlin, Heidelberg},\ \bibinfo {year} {2006})\ pp.\ \bibinfo {pages}
  {144--146}\BibitemShut {NoStop}%
\bibitem [{\citenamefont {{Ole Schuett}}()}]{appmode}%
  \BibitemOpen
  \bibfield  {author} {\bibinfo {author} {\bibnamefont {{Ole Schuett}}},\
  }\href@noop {} {\bibinfo {title} {Appmode: A jupyter extension that turns
  notebooks into web applications}},\ \bibinfo {howpublished} {{G}itHub
  repository: \url{https://github.com/oschuett/appmode}},\ \bibinfo {note}
  {accessed 1 July 2025}\BibitemShut {NoStop}%
\bibitem [{\citenamefont {{IPython Development Team}}()}]{traitlets}%
  \BibitemOpen
  \bibfield  {author} {\bibinfo {author} {\bibnamefont {{IPython Development
  Team}}},\ }\href@noop {} {\bibinfo {title} {Traitlets: Lightweight eventful
  properties for python}},\ \bibinfo {howpublished} {{G}itHub repository:
  \url{https://github.com/ipython/traitlets}},\ \bibinfo {note} {accessed 1
  July 2025}\BibitemShut {NoStop}%
\bibitem [{\citenamefont {{Selenium Development Team}}()}]{selenium}%
  \BibitemOpen
  \bibfield  {author} {\bibinfo {author} {\bibnamefont {{Selenium Development
  Team}}},\ }\href@noop {} {\bibinfo {title} {Selenium browser automation
  project}},\ \bibinfo {howpublished} {\url{https://www.selenium.dev}},\
  \bibinfo {note} {accessed 4 July 2025}\BibitemShut {NoStop}%
\bibitem [{\citenamefont {{Jupyter Development Team}}()}]{jupyterlab}%
  \BibitemOpen
  \bibfield  {author} {\bibinfo {author} {\bibnamefont {{Jupyter Development
  Team}}},\ }\href@noop {} {\bibinfo {title} {Jupyterlab computational
  environment}},\ \bibinfo {howpublished} {{G}itHub repository:
  \url{https://github.com/jupyterlab/jupyterlab}},\ \bibinfo {note} {accessed 4
  July 2025}\BibitemShut {NoStop}%
\bibitem [{\citenamefont {Goscinski}\ \emph {et~al.}()\citenamefont
  {Goscinski}, \citenamefont {Baird}, \citenamefont {Du}, \citenamefont
  {Prado}, \citenamefont {Suman}, \citenamefont {Sodjargal}, \citenamefont
  {Bonella}, \citenamefont {Pizzi},\ and\ \citenamefont
  {Ceriotti}}]{supplementary-code}%
  \BibitemOpen
  \bibfield  {author} {\bibinfo {author} {\bibfnamefont {A.}~\bibnamefont
  {Goscinski}}, \bibinfo {author} {\bibfnamefont {T.~J.}\ \bibnamefont
  {Baird}}, \bibinfo {author} {\bibfnamefont {D.}~\bibnamefont {Du}}, \bibinfo
  {author} {\bibfnamefont {J.}~\bibnamefont {Prado}}, \bibinfo {author}
  {\bibfnamefont {D.}~\bibnamefont {Suman}}, \bibinfo {author} {\bibfnamefont
  {T.-E.}\ \bibnamefont {Sodjargal}}, \bibinfo {author} {\bibfnamefont
  {S.}~\bibnamefont {Bonella}}, \bibinfo {author} {\bibfnamefont
  {G.}~\bibnamefont {Pizzi}},\ and\ \bibinfo {author} {\bibfnamefont
  {M.}~\bibnamefont {Ceriotti}},\ }\href@noop {} {\bibinfo {title}
  {scicode-widgets: Supplementary code for figures}},\ \bibinfo {howpublished}
  {\url{https://doi.org/10.5281/zenodo.15783260}}\BibitemShut {NoStop}%
\bibitem [{iam()}]{iam-notebooks}%
  \BibitemOpen
  \href@noop {} {\bibinfo {title} {Introduction to atomistic modeling, course
  notebooks}},\ \bibinfo {howpublished} {{G}itHub repository:
  {\url{https://github.com/ceriottm/iam-notebooks}}},\ \bibinfo {note}
  {accessed 4 July 2025}\BibitemShut {NoStop}%
\end{thebibliography}
\end{document}